\documentstyle[preprint,aps,floats,epsf]{revtex}
\tightenlines
\draft
\begin{document}
\title{Antikaon Production and Medium Effects
in Heavy Ion Collisions at AGS}
\bigskip
\author{Guang Song$^{1}$, Bao-An Li$^{2}$
\footnote{E-mail: Bali@navajo.astate.edu}, and C. M. Ko$^{1}$
\footnote{E-mail: ko@comp.tamu.edu}}
\address{$^{1}$ Cyclotron Institute and Physics Department,\\
Texas A\&M University, College Station, TX 77843, USA\\
$^{2}$ Department of Chemistry and Physics,\\
Arkansas State University, State University, AR 72467, USA}
\maketitle

\begin{abstract}
Antikaon production from heavy ion collisions at energies available
from the Alternating Gradient Synchrotron (AGS) at the Brookhaven
National Laboratory is studied in a relativistic transport (ART)
model. We include contributions from the baryon-baryon,
meson-baryon, and meson-meson interactions. The final-state
interaction of antikaons via both absorption and elastic scattering
by nucleons and pions are also considered. To compare with
presently available or future experimental data, we have calculated
the antikaon rapidity and transverse momentum distributions as well
as its collective flow. Medium effects on these observables due to
mean field potentials have also been investigated. It is found that
the ratio of antikaon transverse momentum spectrum to that of kaon
and their transverse flow are most sensitive to the in-medium
properties of kaons and antikaons.

\bigskip
\noindent{{\it PACS}: 25.75.-q}

\noindent{{\it Keywords}: Antikaon; Relativistic heavy ion collisions;
Medium effects}
\end{abstract}
\newpage
\section{Introduction}

For heavy ion collisions at the AGS energies, transport model
studies \cite{liko95} have indicated that both density and
temperature of the participant region are high. Heavy ion
experiments at AGS thus offer the possibility to study not only the
hadron to quark-gluon plasma transition but also the properties of
hadrons, such as their masses and lifetimes, in dense medium
\cite{brown2,fang,kl96}. Such medium effects have recently
attracted much attention as they may be related to the precursor
effects due to chiral symmetry restoration \cite{br91,kkl97}. In
particular, knowledge on the properties of kaons in the nuclear
medium is important for understanding both chiral symmetry
restoration and neutron star properties. Since the suggestion by
Kaplan and Nelson on the possibility of kaon condensation in dense
matter \cite{kap86}, there have been many theoretical studies on
the in-medium properties of kaons, using various models, such as
the chiral Lagrangian \cite{lee95}, the boson exchange model
\cite{schaff,schaf97}, the Nambu-Jona-Lasino model \cite{lutz}, and
the coupled-channel approach \cite{koch94,waas96}. These studies
have all shown that $K^+$ feels a weak repulsive potential while
$K^-$ sees a strong attractive potential in the nuclear medium. As
a result, a condensation of antikaons in neutron stars becomes
plausible \cite{kap86,bro87}, which would then lead to the possible
existence of many mini black holes in the galaxies
\cite{bro94,brown}.

A number of experiments have recently been carried out at the AGS
\cite{HIPAGS96}, and preliminary data from these experiments seem
to indicate that medium effects associated with kaons are already
present in some of the observed phenomena \cite{ogilvie}. While the
analysis of experimental data is being finalized, a critical
theoretical examination of both the production mechanism for $K^+$
and $K^-$ and the medium effects on experimental observables will
be very useful. For $K^+$, we have already used the ART model
\cite{liko95} to study its production in heavy ion collisions at
AGS and the dependence of its momentum spectra on its in-medium
properties \cite{liko96}. In particular, we have found that the
$K^+$ transverse collective flow is sensitive to the kaon
dispersion relation in dense nuclear matter. In the present work,
we shall report a similar study for $K^-$.

In Section II, we will briefly describe the ART model used in our
previous study of particle production and signatures of chiral
symmetry restoration and/or QGP formation in heavy ion collisions
at AGS energies \cite{liko95,liko96,liko96b}. We will then discuss
the details of implementing various reaction channels for $K^-$
production. These include $K^-$ production from baryon-baryon,
meson-baryon, and meson-meson interactions. Also, $K^-$ absorption
and its final-state elastic scattering will be considered. In
Section III, results from this study are presented. In particular,
we shall discuss the relative contributions of different reaction
channels to the $K^-$ yield, its rapidity and transverse mass
distribution and collective flow. Medium effects due to both the
nuclear and Coulomb potentials on these observables will also be
investigated. The beam energy dependence of the medium effects on
both $K^+$ and $K^-$ will be studied. Finally, a summary is given
in Section IV.

\section{Antikaon production, absorption and rescattering in the ART model}

The ART model is a pure hadronic transport model developed for
modeling relativistic heavy ion collisions up to the AGS energies
\cite{liko95}. For completeness, we summarize here the main
features of this model and refer the reader to Ref. \cite{liko95}
for its details. The ART 1.0 includes the following baryons:
$N,~\Delta(1232),~N^{*}(1440),~N^{*}(1535), ~\Lambda,~\Sigma$; and
mesons: $\pi,~K,~\eta,~\rho,~\omega$; as well as their explicit
charge states. Both elastic and inelastic collisions among most of
these particles are modeled as best as we can by using as inputs
the experimental data from hadron-hadron collisions. Most inelastic
hadron-hadron collisions are modeled through the formation of
baryon and meson resonances. Although we have only explicitly
included three baryon resonances, effects of heavier baryon
resonances with masses up to 2 GeV are partially taken into account
through the formation of these resonances in the intermediate
states of meson-baryon reactions. We have also included in the
model optional, self-consistent mean field potentials for both
baryons and kaons.

The treatment of antikaon in the ART model 1.0 is, however,
incomplete as only antikaon production from meson-meson
interactions has been included. Although this has negligible
effects on the reaction dynamics and experimental observables
associated with nucleons, pions and kaons, we have been unable to
study in detail the production mechanism for antikaon and its
dependence on the medium effects. In this section, we shall first
discuss the improvement we have made in the ART model for treating
antikaon production, absorption, rescattering, and propagation.

\subsection{Antikaon production from baryon-baryon interactions}

There are few experimental data on $K^{-}$ production from
nucleon-nucleon interactions in the energy range we are
considering. In particular, its total production cross section from
$pp$ collisions is practically unknown. There are a number of
parameterizations of the antikaon production cross section from the
nucleon-nucleon interaction \cite{efr94,sib97,li97,zwer84}. For
example, the inclusive $K^{-}$ production cross section from the
proton-proton interaction, i.e., $pp \to K^{-}X$, has been
parameterized in Ref. \cite{efr94} using phase space consideration.
We choose this one in the present work as it fits better the
available data at high energies. Specifically, the $K^-$ production
cross section from $pp$ collisions is given by
\begin{equation}
\sigma_{pp \to K^-X}(s) = \left (1-{{s_{\rm 0}}\over{s}}\right )^3
\left [2.8F_1\left ({{s}\over{s_{\rm 0}}}\right )+
7.7F_2\left ({{s}\over{s_{\rm 0}}}\right )\right ]+
3.9F_3\left ({{s}\over{s_{\rm 0}}}\right ) \  [{\rm mb}],
\end{equation}
with
\begin{eqnarray}
F_1(x)&=&(1+1/\sqrt{x}){\rm ln}(x)-4(1-1/\sqrt{x}),\nonumber\\
F_2(x)&=&1-(1/\sqrt{x})(1+{\rm ln}(x)/2),\nonumber\\
F_3(x)&=&\left ({{x-1}\over {x^2}}\right )^{3.5}.\nonumber
\end{eqnarray}
In the above,
${s_{\rm 0}}^{1/2}$ = 2($m_p+m_K$)=2.8639\ GeV is the threshold energy.

At AGS energies the final state is expected to be dominated by two
nucleons and a kaon-antikaon pair. We thus assume that the cross
section for $pp\to ppK^+K^-$ is the same as that for $pp\to K^-X$.
Since there are no data for $K^-$ production from $np$, $nn$ and
other baryon-baryon interactions involving one or two resonances,
to determine their cross sections thus requires models for these
interactions. In the present study, we make the minimum assumption
that they all have the same $K^-$ production cross section as in
$pp$ collisions at the same center of mass energy.

To determine the momentum distribution of the produced $K^-$, we
make use of the empirical observation that the momentum
distributions of final particles in high energy $pp$ collisions all
have the following form
\begin{equation}\label{dis}
\frac{d^2\sigma}{dp_T^2dp_L}\propto e^{-A{x^*}^2}e^{-Bp_T^2},
\end{equation}
where $x^*=p_{\rm L}/p_{\rm L_{max}}$ with $p_{\rm L_{max}}$ being
the maximum longitudinal momentum of the particle. In the ART
model, this is carried out by first obtaining the $K^-$ and $K^+$
transverse and longitudinal momenta from the above distribution,
assuming that the angle between $p_x$ and $p_y$ is uniformly
distributed. Then, the longitudinal momenta of the two baryons are
obtained using also a similar distribution. From energy and
momentum conservation, the transverse momenta of both baryons can
be determined. We find that the limited experimental data
\cite{diddens} are reasonably fitted by using the following values:
$A=12.5$ and $B=4.15$ (GeV/c)$^2$ for $K^-$; $A=5.3$ and $B=3.68$
(GeV/c)$^2$ for $K^+$; and $A=2.76$ (GeV/c)$^2$ for $N$.

\subsection{Antikaon production from meson-baryon interactions}

The cross section for $K^-$ production from pion-nucleon
interactions has been studied in Ref. \cite{sib97} using a
boson-exchange model. Reactions with one or more pions in the final
state are neglected as their cross sections are small in the energy
range we consider. Following Ref. \cite{sib97}, we have
\begin{eqnarray}
2\sigma(\pi^{-}p\to p K^{0} K^{-})&=&
\sigma(\pi^{-}p\to n K^{+} K^{-})=
\sigma(\pi^{-}n\to n K^{0} K^{-})\nonumber\\
=4\sigma(\pi^{0}p\to p K^{+} K^{-})&=&
4\sigma(\pi^{0}n\to n K^{+} K^{-})=
\sigma(\pi^{0}n \to p K^{0} K^{-})\nonumber\\
=\sigma(\pi^{+}n \to p K^{+} K^{-})&=&
\sigma_0,
\end{eqnarray}
where $\sigma_0$ is given by \cite{sib97}
\begin{equation}
\sigma_0 = 1.21(1-s_0/s)^{1.86}(s_0/s)^2 \ [{\rm mb}].
\end{equation}

For $\pi$-baryon resonance and $\rho (\omega)$-baryon collisions
there are no experimental data. We again make the minimum
assumption that their cross sections are the same as that in
pion-nucleon interaction at the same center of mass energy.

The momentum distribution of the produced $K^-$ from meson-baryon
interactions is determined by the three-body phase space.

\subsection{Antikaon production from meson-meson interactions}

As in the original ART model \cite{liko95}, antikaon productions
from all meson-meson collisions are modeled through the process $MM
\to K\bar{K}$. The cross section used for $\pi\pi \to K\bar{K}$ is
calculated from the $K^*$ exchange model of Ref. \cite{brown2}. All
other meson-meson reactions, such as $\rho\rho \to K\bar{K}$ or
$\pi\rho\to K\bar{K}$, are taken to have a constant value of 0.3
mb. Contrary to antikaon production from baryon-baryon and
meson-baryon interactions, the momentum of produced antikaon from
meson-meson interaction is trivially fixed by kinematics as the
final state consists of only two particles.

\subsection{Antikaon absorption and its production from meson-hyperon
interactions}

Antikaons produced in hot dense matter can be absorbed by nucleons
via strange-exchange reactions. For final states consisting of a
$\Sigma$ particle, we have the following reactions:
\begin{eqnarray}
   K^{-}p&\to& \Sigma^{0}\pi^{0}, \Sigma^{-}\pi^{+},
   \Sigma^{+}\pi^{-},\nonumber\\
   K^{-}n&\to& \Sigma^{0}\pi^{-}, \Sigma^{-}\pi^{0}.
\end{eqnarray}
Their cross sections are taken from Ref. \cite{cug90}, i.e.,
\begin{eqnarray}
\sigma(K^-p \to \Sigma^0\pi^0)& = &0.6p^{-1.8}\ [{\rm mb}];\ 0.2
\leq p \leq 1.5 {\rm GeV/c},\nonumber\\
\sigma(K^-n \to \Sigma^0\pi^-)& = & \left\{ \begin{array}{ll}
1.2p^{-1.3} \ [{\rm mb}], &\mbox{${\rm if} \ 0.2 \leq p \leq 1 {\rm
GeV/c}$;}\\
1.2p^{-2.3}\ [{\rm mb}], &\mbox{${\rm if} \ 1 \leq p \leq 6
{\rm Gev/c}$.}  \end{array}
\right.
\end{eqnarray}
where $p$ is the $K^-$ momentum in the laboratory frame.

For $K^-$ absorption by resonances, we take the cross sections for
$N^{*+}$ and $\Delta^{+}$ to be the same as for $p$, while those
for $N^{*0}$ and $\Delta^0$ the same as for $n$. We also include
the reactions $K^{-}\Delta^{++}$ $\to$ $\Sigma^{+}\pi^{0}$,
$\Sigma^{0}\pi^{+}$ by assuming that their cross sections are the
same as that for $K^{-}p$. On the other hand, the cross section for
$K^{-}\Delta^{-}$ $\to$ $\Sigma^{-}\pi^{-}$ is taken to be the same
as that for $K^{-}n$.

For final states with a $\Lambda$, the following reactions are possible,
\begin{eqnarray}
K^{-}p &\to& \Lambda + \pi^{0}, \nonumber\\
         K^{-}n &\to& \Lambda + \pi^{-}.
\end{eqnarray}
The $K^-$p cross section is parameterized as in Ref. \cite{cug90}, i.e.,
\begin{equation}
\sigma_{K^-p\to \Lambda\pi^0} = \left\{ \begin{array}{ll}
50p^2-67p+24\ [{\rm mb}], &\mbox{if \ $0.2 \leq p \leq 0.9 GeV$;} \\
3p^{-2.6}\ [{\rm mb}], &\mbox{if \ $0.9 \leq p \leq 10 GeV$.}
\end{array}
\right.
\end{equation}
For other cross sections involving $n$ and baryon resonances, they
are assumed to be the same as $K^-p\to\Lambda\pi^0$.

In the inverse reactions, i.e. $\pi+\Lambda(\Sigma)$, antikaon can
be reproduced, and their cross sections are deduced from the
antikaon absorption cross section using the detailed balance
relations. A hyperon can also interact with other mesons, such as
rho and omega to produce a $K^-$. The corresponding cross sections
are assumed to be the same as those for a pion at the same center
of mass energy.

\subsection{Antikaon final-state interactions}

Because of final-state interactions, not all antikaons produced in
hadron-hadron interactions can escape freely from the reaction zone
in heavy ion collisions. The absorption reactions have already been
described in the previous section. Here, we are mainly concerned
with the $K^-$ elastic scatterings with baryons. These cross
sections are taken from Ref. \cite{cug90}, i.e.,
\begin{eqnarray}
\sigma_{K^-p\to K^-p}& = &13p^{-0.9}\ [{\rm  mb}], \ 0.25
\leq p \leq 4.0\ {\rm GeV/c}. \nonumber\\
\sigma_{K^-n\to K^-n}& = & \left\{ \begin{array}{ll}
20.0p^{-2.74}\ [{\rm mb}], & \mbox{if $0.5 \leq p \leq 1.0$ GeV/c;} \\
20.0p^{-1.8}\ [{\rm mb}],  & \mbox{if $1.0 < p \leq 4.0$ GeV/c.}
\end{array}
\right.
\end{eqnarray}

\subsection{Mean field potentials for kaons and antikaons}

In transport models, the imaginary part of the self energy of a
hadron is approximately treated by its scatterings with other
hadrons, while the real part of the self energy is given by the
mean field potential. Various approaches have been used to evaluate
the kaon mean field potential in the nuclear medium
\cite{lee95,schaff,schaf97,lutz,koch94,waas96}, we use here the one
determined from the kaon-nucleon scattering length $a_{KN}$ in the
impulse approximation \cite{brown}, i.e.,
\begin{equation}
\omega(p,\rho_b)=\left[m_K^2+p^2-4\pi\left(1+\frac{m_K}{m_N}\right)
a_{KN}\rho_b\right]^{1/2},
\end{equation}
where $m_K$ and $m_N$ are the kaon and nucleon masses,
respectively; $p$ is the kaon momentum; $\rho_b$ is the baryon
density; and $a_{KN}\approx -0.255$ fm is the isospin-averaged
kaon-nucleon scattering length. The kaon potential in the nuclear
medium is given by
\begin{equation}
U(p,\rho_b)=\omega(p,\rho_b)-(m_K^2+p^2)^{1/2}.
\end{equation}
At normal nuclear density, a kaon at rest has a repulsive potential
of about 30 MeV.

For the $K^-$ potential, we use the following expression
\begin{equation}
 U(\rho)=-0.12(\rho_b/\rho_0)~[{\rm GeV}].
\end{equation}
The magnitude of this potential is similar to that either extracted
from the experimental data on kaonic atom or predicted by
theoretical models \cite{lee95,schaff,schaf97,lutz,koch94,waas96}.

One of the main purposes of this work is to explore the effects of
kaon potentials on experimental observables. We shall thus compare
in the next section results from calculations with and without mean
field potentials for kaons and antikaons. Although the model at the
present stage probably cannot distinguish the different forms of
kaon and antikaon potentials from various theories, it does predict
that some observables are appreciably affected by the presence of
mean field potentials as shown below.

\section{Results and discussions}

We now turn to the results of our study. After including all $K^-$
production channels outlined above, our previous predictions, that
are based on the ART 1.0, for the nucleon, pion and kaon
observables do not show significant changes as antikaons are small
perturbations to the collision dynamics and other hadrons. We thus
only present in the following results for antikaons.

\subsection{Production mechanism for antikaons}

\begin{figure}[htp]
\setlength{\epsfxsize=3.5truein}
\centerline{
\vskip 0.2truein
\hskip 1.25truein
\epsffile{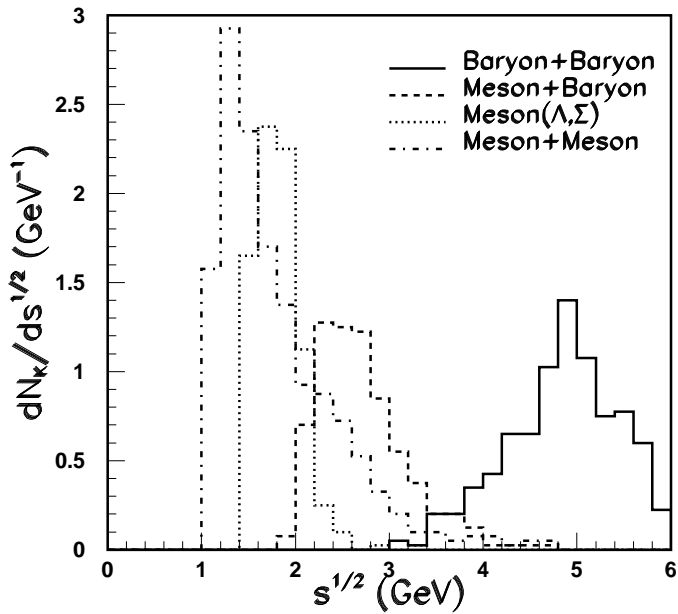}
\vskip 0.3truein
\caption{Center of mass energy distribution of hadron pairs with
energies above the $K^-$ production threshold in the reaction of Au
+ Au at $p_{\rm beam}$/A = 11.6 GeV/c and impact parameter b = 4
fm, and with the $K^-$ mean field potential.}}
\end{figure}

To identify the sources for antikaon production, we show in Fig. 1
the center of mass energy distribution of hadron pairs with
energies above the $K^-$ production threshold in the reaction of Au
+ Au at $p_{\rm beam}$/A = 11.6 GeV/c and impact parameter of 4 fm.
It is seen that contributions from meson-meson and meson-baryon
collisions are more important than that from baryon-baryon
collisions. More quantitatively, about 40\%, 40\% and 20\% of
produced $K^-$ are from meson-meson, meson-baryon, and
baryon-baryon collisions, respectively. The relative importance of
different $K^-$ sources seen here thus does not agree completely
with results from either RQMD \cite{rqmd}, where meson-baryon
interactions seem to contribute most, or the ARC \cite{arc} model,
which shows a main contribution from baryon-baryon interactions. We
attribute the different conclusions to the fact that different
assumptions about the cross sections for the elementary $K^-$
production reactions are introduced in these models. Similar
differences have also been seen previously in transport model
studies of $K^+$ production \cite{liko95}.

To learn about the dynamics of antikaon production, the primordial
$K^-$ multiplicity is shown in Fig. 2 as a function of time for
various $K^-$ production channels by turning off the $K^-$
absorption reactions in the calculation. It is seen that $K^-$
production starts at about t=0.5 fm/c after the contact of the
colliding nuclei and is initially dominated by the contribution
from baryon-baryon collisions. Soon after that, meson-baryon
collisions begin to contribute significantly. The meson-meson
collisions do not contribute until around 2 fm/c and become the
dominant ones after 4 fm/c. Both $K^-$ production and absorption
from meson-meson collisions and their inverse reactions are seen to
last longer than all other reactions. We note that $K^-$ production
from baryon-baryon collisions practically ceases after about 4
fm/c.

\begin{figure}[htp]
\setlength{\epsfxsize=3.5truein}
\centerline{
\vskip 0.2truein
\hskip 1.25truein
\epsffile{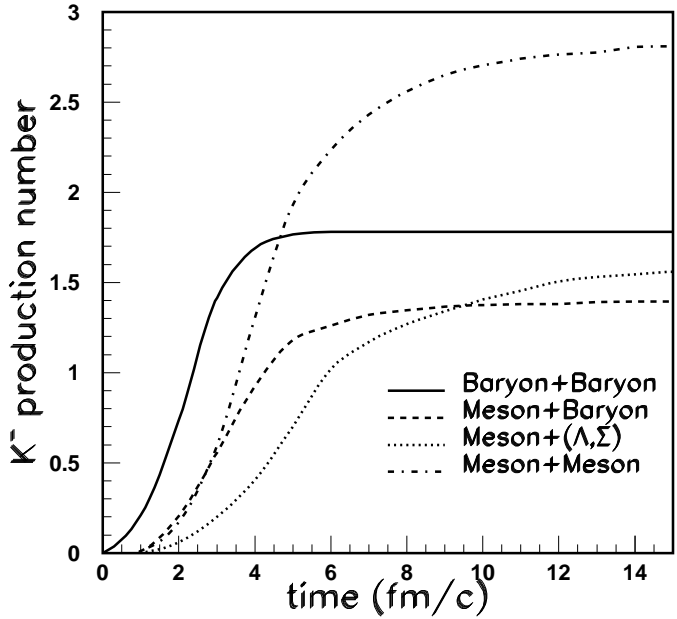}
\vskip 0.3truein
\caption{Primordial $K^-$ number from
different reaction channels as functions of time in the same
reaction as in Fig. 1.}}
\end{figure}

Including $K^-$ absorption we have repeated the calculations for
the same reaction as in Fig. 2, and the results are shown in Fig.
3. Comparing Fig. 2 and Fig. 3 one sees that close to half of the
$K^-$'s produced from baryon-baryon collisions are absorbed during
their propagation through the system. For both meson-baryon and
meson-hyperon collisions, about 40\% of the primordial $K^-$ are
absorbed, while it is only about one fourth for those created
through meson-meson collisions. This is mainly because $K^-$'s from
baryon-baryon collisions are produced earlier in the collisions, so
they spend a longer time in the dense matter and thus have a higher
chance to get absorbed. Another reason is that $K^-$'s from
baryon-baryon collisions generally have higher kinetic energies and
thus have velocities comparable to those of baryons, so it is
easier for them to be absorbed.

\begin{figure}[htp]
\setlength{\epsfxsize=3.5truein}
\centerline{
\vskip 0.2truein
\hskip 1.25truein
\epsffile{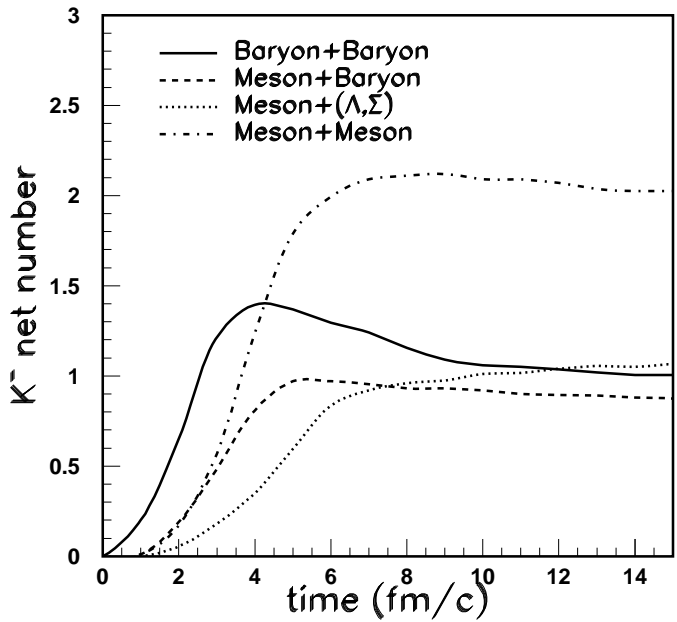}
\vskip 0.3truein
\caption{Net $K^-$ number after absorption for the same reaction as
in Fig. 1.}}
\end{figure}

\subsection{Mean field effects on antikaon spectra and yields}

To study the effects of mean field potentials on kaons and
antikaons, we have calculated their transverse mass spectra and
rapidity distributions with and without mean field potentials. We
shall also compare them with recent data from the E802/E866
collaboration \cite{ags93}.

Fig. 4 shows the $K^+$ and $K^-$ transverse mass distributions from
the reaction of Au + Au at $p_{\rm beam}$/A = 11.6 GeV/c and impact
parameters $b \leq$ 4 fm. It is seen that the attractive mean field
potential pulls $K^-$ to lower values of transverse momentum,
causing its slope to increase, while for $K^+$, the effect seems to
be opposite. Since the mean field potential is stronger for $K^-$
than for $K^+$, the effect is also larger for $K^-$ than for $K^+$.
While the $K^+$ data can be reproduced reasonably well by both
calculations with and without the mean field potential, the
calculations with the mean field potential for $K^-$ seem to better
reproduce the data.

\begin{figure}[htp]
\setlength{\epsfxsize=3.5truein}
\centerline{
\vskip 0.2truein
\hskip 1.25truein
\epsffile{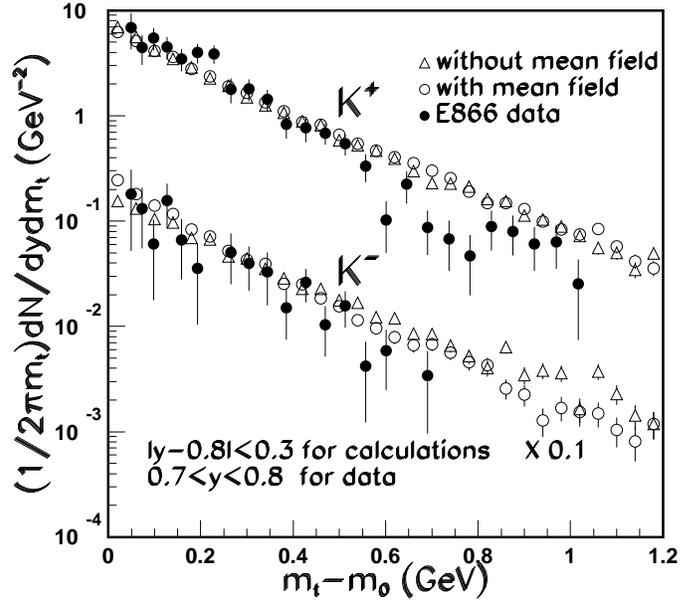}
\vskip 0.3truein
\caption{Transverse mass spectra of $K^+$ and $K^-$ in the reaction
of Au + Au at $p_{\rm beam}$/A = 11.6 GeV/c and impact parameters
$b \leq$ 4 fm.}}
\end{figure}

It has been proposed that the $K^+/K^-$ ratio as a function of
their transverse mass is a sensitive probe of mean field effects
\cite{liko96}. This is because the average thermal velocities of
$K^+$ and $K^-$, which are much larger than the change of velocity
caused by medium effects, are almost canceled out. Indeed, as shown
in Fig. 5 for the same reaction as in Fig. 4 this ratio decreases
with the transverse mass $m_{\rm t}$ in the case without mean field
potentials but increases with $m_{\rm t}$ once mean field
potentials are included.

\begin{figure}[htp]
\setlength{\epsfxsize=3.5truein}
\centerline{
\vskip 0.2truein
\hskip 1.25truein
\epsffile{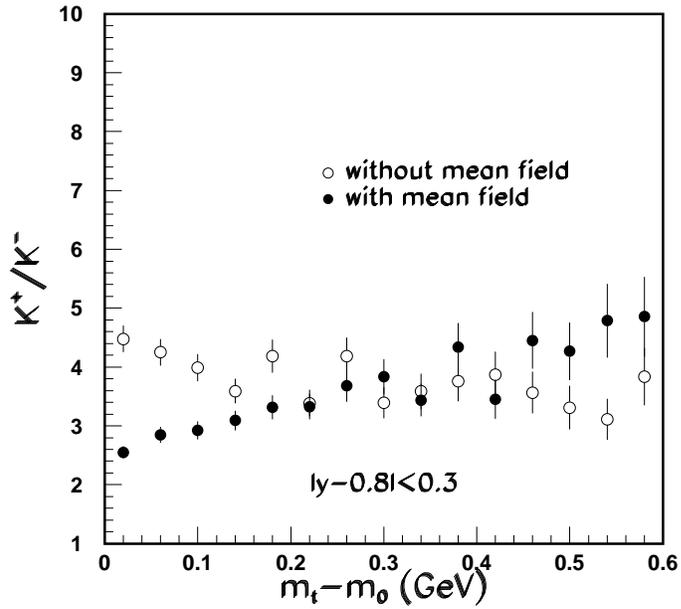}
\vskip 0.3truein
\caption{Transverse mass spectra of $K^+/K^-$ in the same reaction as
in Fig. 4.}}
\end{figure}

\begin{figure}[htp]
\setlength{\epsfxsize=3.5truein}
\centerline{
\vskip 0.2truein
\hskip 1.25truein
\epsffile{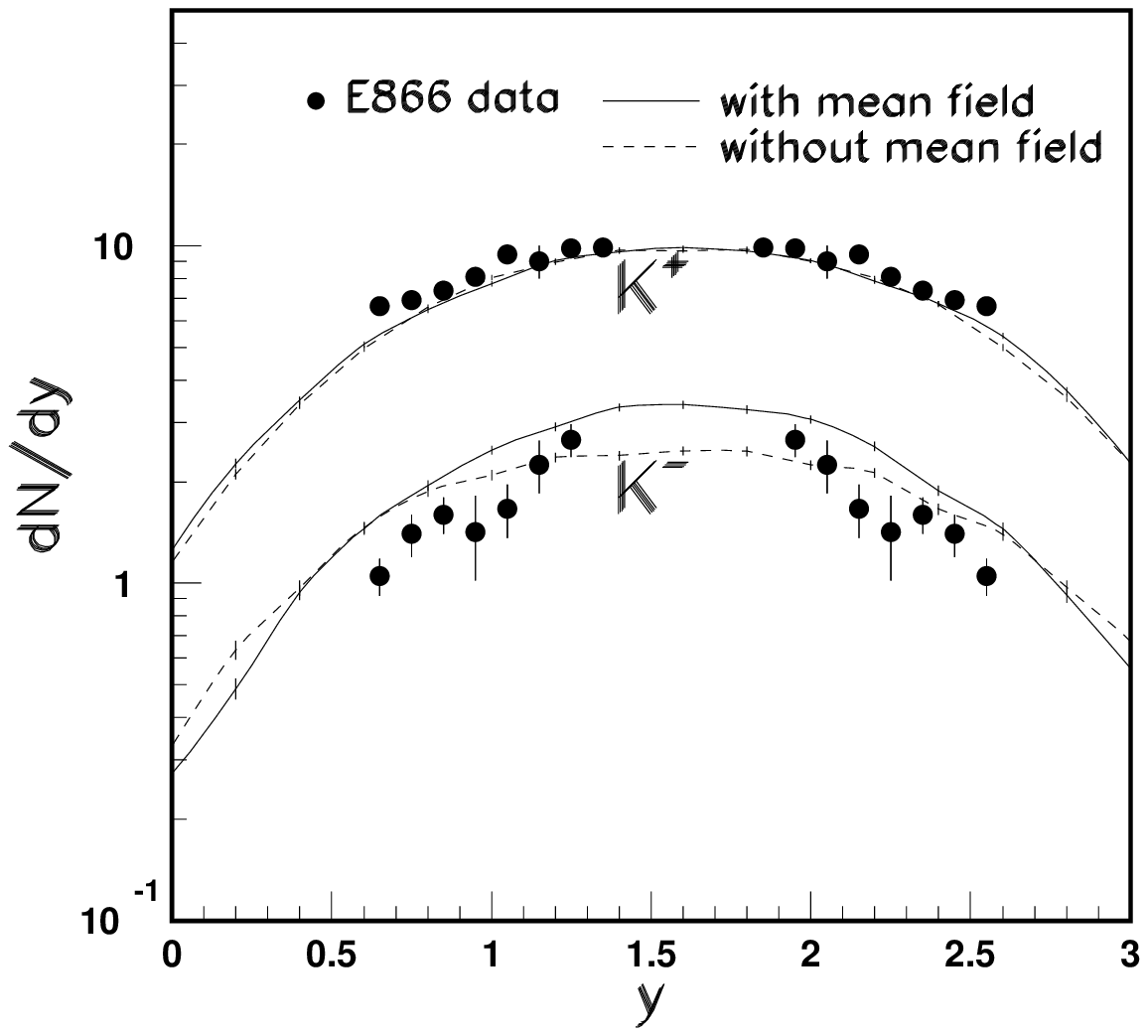}
\vskip 0.3truein
\caption{Rapidity distributions of $K^+$ and $K^-$ in the same reaction
as in Fig. 4.}}
\end{figure}

The $K^+$ and $K^-$ rapidity distributions for the same Au+Au
reaction as in Fig. 4 are shown in Fig. 6. The rapidity
distribution for $K^-$ is seen to be narrower than that for $K^+$.
This may also be a signature of the mean field effect as the strong
attractive $K^-$ mean field potential makes $K^-$ less energetic,
leading thus to a narrower rapidity distribution around the
mid-rapidity than in the case without a potential. Although these
low energy $K^-$'s at the central rapidity are more easily absorbed
by nucleons, the inverse reactions of $K^-$ production from
pion-hyperon interactions is also more important as most pions and
hyperons are concentrated at the central rapidity. One thus expects
an increase of the $K^-$ yield after including the mean field
potential. The theoretical results support such a picture. Indeed,
without mean field potentials the rapidity distribution for $K^+$
is narrower than that for $K^-$ but becomes much wider after the
mean field potentials are included. Also, the increase in the $K^-$
yield at central rapidity is more than its decrease at the
projectile and target rapidities, leading to an increased total
$K^-$ yield when the mean field potential is introduced. This
effect can be more clearly seen in Fig. 7 where the ratio $K^+/K^-$
is shown as a function of rapidity $y$. It shows that the ratio
decreases from about 4.2 around the mid-rapidity to about 3.5
around the target and projectile rapidities if no kaon mean field
is included. Medium effects change this ratio dramatically; it
increases from about 3.0 at the mid-rapidity to about 4.0 at the
target and projectile rapidities.

\begin{figure}[htp]
\setlength{\epsfxsize=3.5truein}
\centerline{
\vskip 0.2truein
\hskip 1.25truein
\epsffile{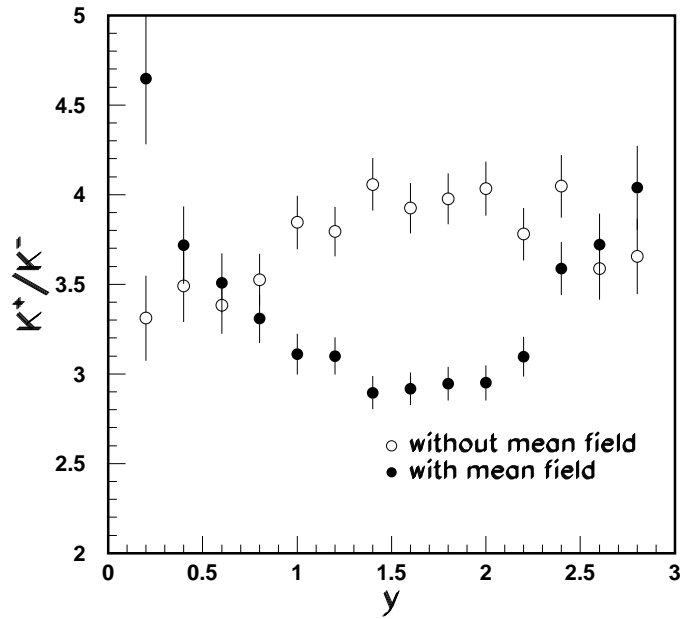}
\vskip 0.3truein
\caption{Rapidity distributions of the $K^+/K^-$ ratio in
the same reaction as in Fig. 4.}}
\end{figure}

We note that both $K^+$ and $K^-$ production from Au+Au collisions
at $p_{\rm beam}$/A = 11.6 GeV/c have also been studied in the HSD
model \cite{cassing}, which includes both initial string dynamics
and subsequent hadron cascade. Contrary to our results as well as
those from the ARC and RQMD models, the HSD model underpredictes
the yield of both $K^+$ and $K^-$. This may be due to the
introduction in the model a finite formation time, which is not
included in either our model or the ARC model. Although finite
formation time is included in the RQMD model, the resulting large
yield of $K^+$ and $K^-$ is probably due to the inclusion of high
mass resonances, which are neglected in the HSD model. More work is
thus required to clarify the effects due to different physical
assumptions in these transport models.

\subsection{Transverse flow analysis for antikaons}

It was first demonstrated in Ref. \cite{lkl95} that kaon transverse
flow is a powerful probe of kaon in-medium potentials in heavy ion
collisions at SIS/GSI energies, which are an order of magnitude
lower than that at AGS. Subsequently, using ART 1.0 the kaon
transverse flow has also been found to be the most sensitive
observable for studying the kaon dispersion relation in the dense
medium formed in relativistic heavy ion collisions at AGS energies
\cite{liko96}. It is thus interesting to compare the transverse
flow of kaons and antikaons as their mean field potentials have
opposite signs.

First, we perform the standard transverse flow analysis for $K^-$
for the same Au+Au reaction as in Fig. 4, and the results are shown
in Fig. 8. It is seen that without $K^-$ potential (shown by open
circles) antikaons flow in the opposite direction as that of
nucleons since antikaons flowing with nucleons are absorbed due to
strong strange-exchange reactions. Such a shadowing effect due to
the spectator matters has also been seen in the transverse flow of
pions \cite{gos89,bal91,bal94,kin97}. However, when the attractive
mean field potential is included (shown by solid circles), the flow
of antikaons change its direction toward that of nucleons.

\begin{figure}[htp]
\setlength{\epsfxsize=3.5truein}
\centerline{
\vskip 0.2truein
\hskip 1.25truein
\epsffile{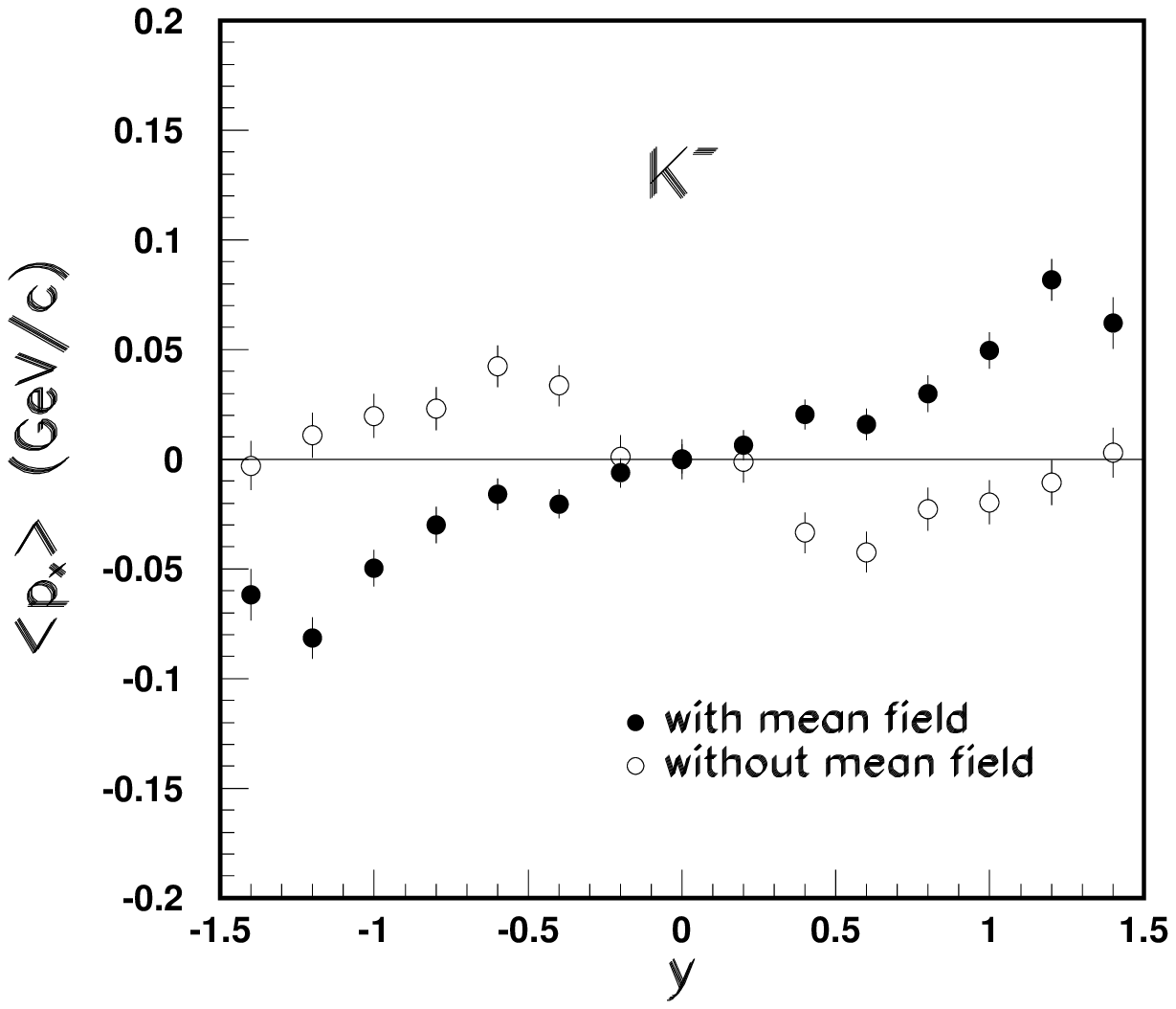}
\vskip 0.3truein
\caption{Transverse flow of $K^-$ in the same reaction as in Fig. 4.}}
\end{figure}

For comparison we show in Fig. 9 the transverse flow of $K^+$
calculated with and without mean field potentials for the same
Au+Au reaction as in Fig. 8. For both $K^+$ and $K^-$, their
transverse flow are reversed once mean field potentials are
introduced in the transport model. The transverse flow of antikaons
is, however, found to be more sensitive to the mean field potential
due to the stronger antikaon potential compared to that for kaon.
This makes antikaon flow analysis an even more valuable tool for
studying the in-medium properties of antikaons. It is interesting
to mention that the preliminary data from the E866 collaboration on
the $K^+$ and $K^-$ flow is consistent with those predicted by an
attractive mean field potential for $K^-$ and a repulsive one for
$K^+$ \cite{ogilvie97}.

\begin{figure}[htp]
\setlength{\epsfxsize=3.5truein}
\centerline{
\vskip 0.2truein
\hskip 1.25truein
\epsffile{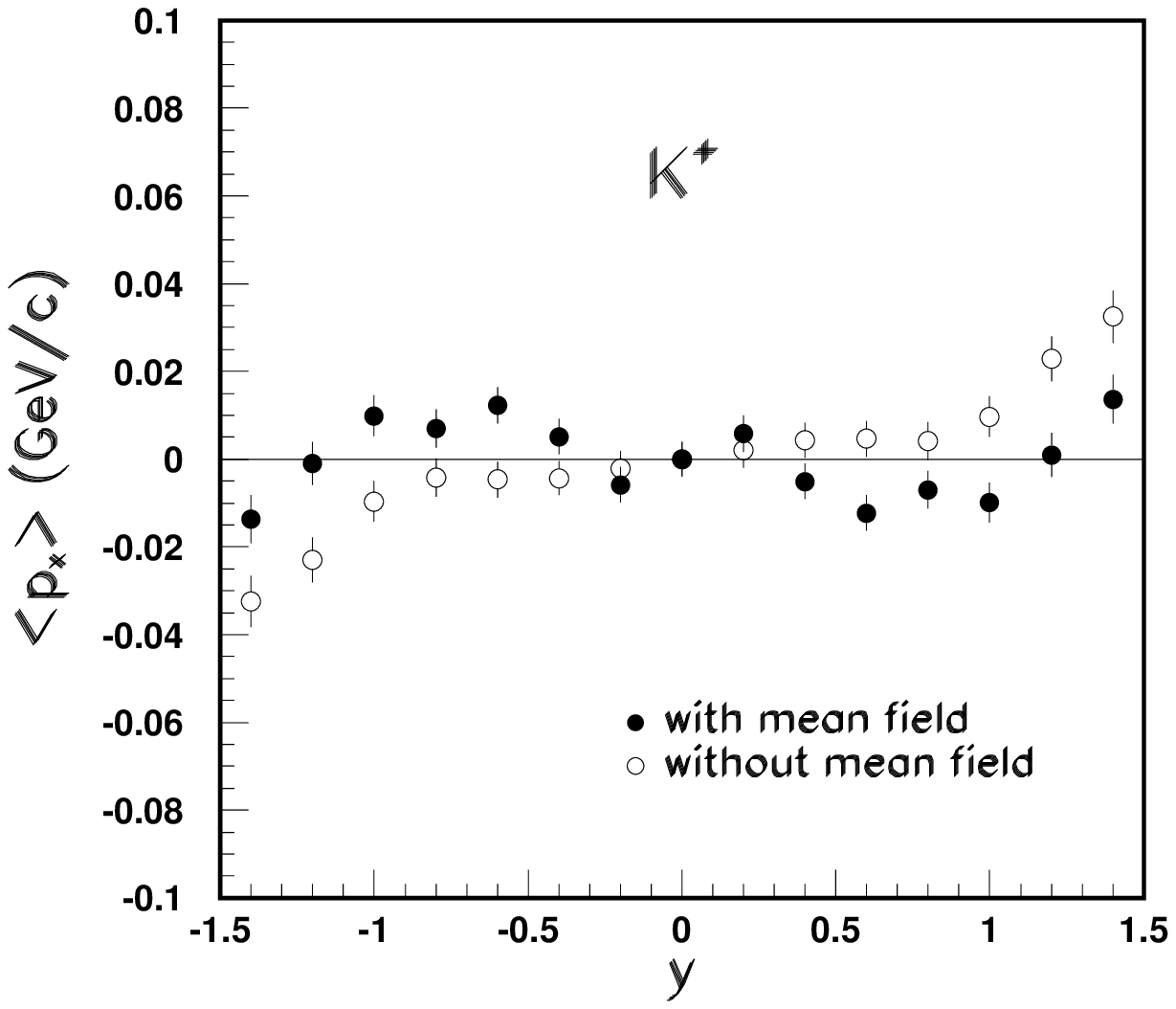}
\vskip 0.3truein
\caption{Transverse flow of $K^+$ in the same reaction as in
Fig. 4.}}
\end{figure}

\subsection{Coulomb effects on the transverse momentum
spectra of kaons and antikaons}

Besides the mean field potential due to strong interaction, both
$K^+$ and $K^-$ are also affected by their Coulomb potentials.
While the mean field potential is attractive and repulsive for
$K^-$ and $K^+$, respectively, the Coulomb potentials have similar
effects. To see the relative importance of mean field and Coulomb
potentials on the spectra of $K^+$ and $K^-$, the following four
different calculations have been carried out: with Coulomb
potential only, with mean field potential only, with both
potentials, and without any potential. Results of these studies are
shown in Fig. 10 for the same reaction as in Fig. 4. It is seen
that for $K^-$ the effect due to the mean field potential is much
stronger than that due to the Coulomb potential. This is different
from the $K^+$ case, where the mean field potential is weaker than
that for $K^-$, and its effects is thus comparable to that due to
the Coulomb potential. We conclude that mean field effects on the
$K^-$ transverse mass spectra, especially at lower masses, are
distinguishable from that due to the Coulomb potential.

\begin{figure}[htp]
\setlength{\epsfxsize=3.5truein}
\centerline{
\vskip 0.2truein
\hskip 1.25truein
\epsffile{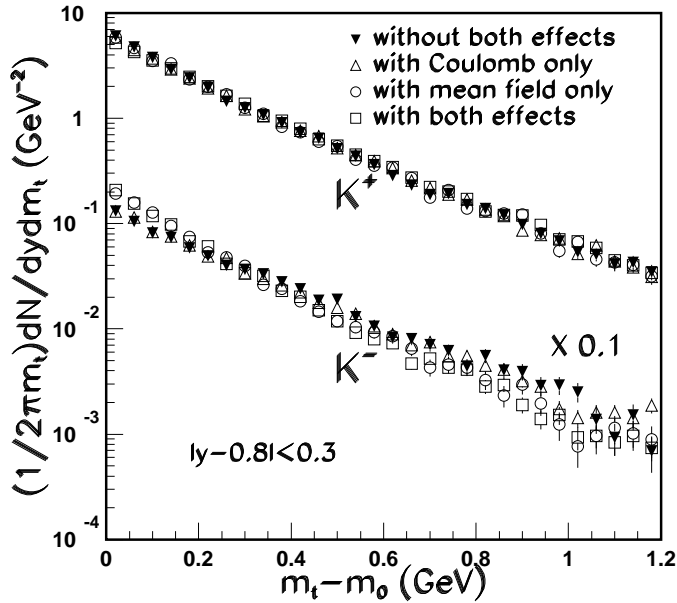}
\vskip 0.3truein
\caption{Coulomb and nuclear mean field effects on transverse
mass distributions of $K^+$ and $K^-$ in the same reaction as in
Fig. 4.}}
\end{figure}

\subsection{Beam energy dependence of the mean field effects on antikaons}

Heavy ion collision dynamics is governed by both individual
hadron-hadron collisions and mean field potentials. For baryons,
mean field effects have been found to be more important in
collisions at lower energies. However, it is not clear what is the
beam energy dependence of the mean field effects on produced
particles. To answer this question we have carried out an analysis
of the rapidity distribution and transverse mass spectrum of both
kaons and antikaons in the beam energy range of 2 to 16 GeV/nucleon
with and without mean field potentials. In this section, the beam
energy dependence of mean field effects on antikaons is presented.

Fig. 11 shows the $K^-$ transverse mass spectra calculated with and
without the mean field potential for Au+Au reactions at an impact
parameter of 4 fm and a beam energy of 4, 10.7 and 16 GeV/nucleon,
respectively. It is seen that the mean field effect is almost the
same at all three beam energies. To be more quantitative the $K^-$
inverse slope has been extracted by fitting the $m_{\rm t}$ spectra
with exponential functions. We note that this parameter should not
be identified as the temperature as effects due to collective
radial flow have not been corrected. In Fig. 12, the inverse slope
parameter of the $K^-$ transverse mass spectrum is shown as a
function of beam energy. It is seen that it increases with beam
energy in both cases as one would expect. Moreover, the mean field
potential reduces the $K^-$ inverse slope parameter by about 20\%
in the whole beam energy range.

\begin{figure}[htp]
\setlength{\epsfxsize=3.5truein}
\centerline{
\vskip 0.2truein
\hskip 1.25truein
\epsffile{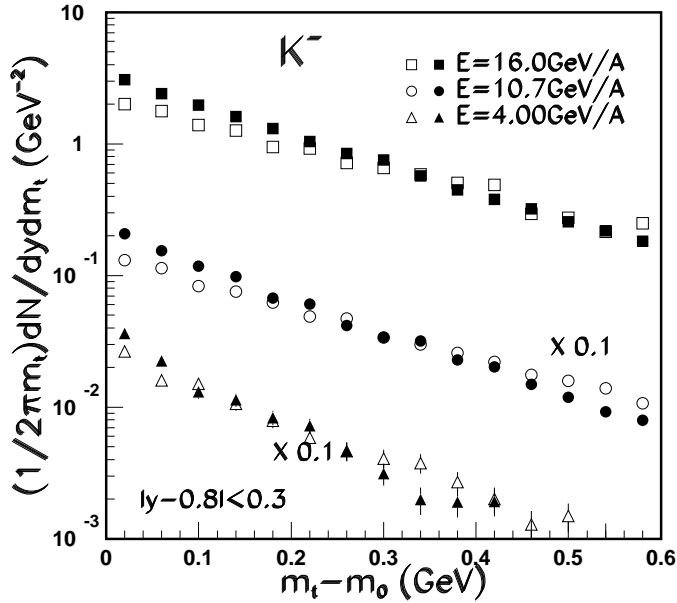}
\vskip 0.3truein
\caption{Transverse mass spectra of $K^-$ with and without the
$K^-$ mean field potential in the reaction of Au + Au at different
beam energies and an impact parameter $b$ = 4 fm. Solid and open
circles are, respectively, for the case with and without mean field
potentials.}}
\end{figure}

\begin{figure}[htp]
\setlength{\epsfxsize=3.5truein}
\centerline{
\vskip 0.2truein
\hskip 1.25truein
\epsffile{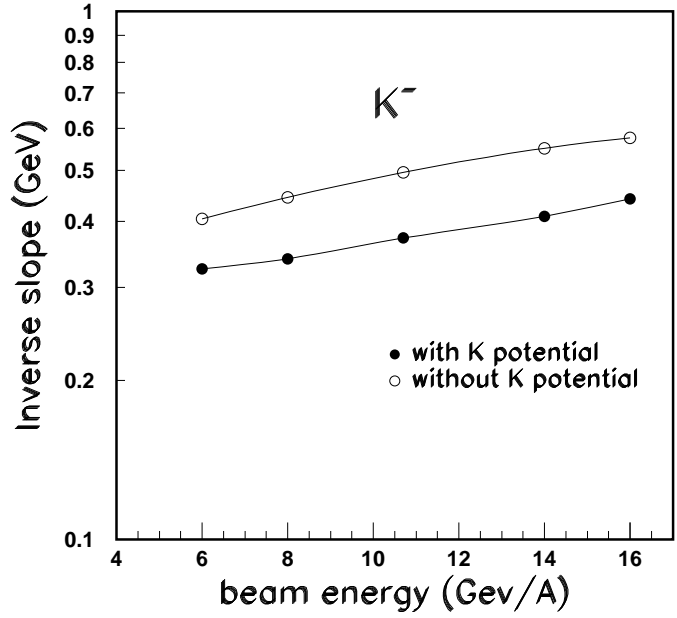}
\vskip 0.3truein
\caption{Inverse slope or apparent temperature of $K^-$
with and without $K^-$ mean field potential in collisions of Au +
Au at different beam energies and impact parameter $b$ = 4 fm.}}
\end{figure}

The rapidity distributions of antikaons from the same reaction as
in Fig. 11 are compared in Fig. 13. It is again seen that the mean
field effect on $K^-$ rapidity distribution changes very little as
the beam energy varies. We note that a similar observation has also
been found for kaons \cite{liko95}. We thus conclude that the mean
field effects on kaons and antikaons are essentially independent of
the beam energy in the energy range of 2 to 16 GeV/nucleon. Several
experimental collaborations are currently studying the beam energy
dependence of particle production at the AGS, so our results can be
tested in the near future.

\begin{figure}[htp]
\setlength{\epsfxsize=3.5truein}
\centerline{
\vskip 0.2truein
\hskip 1.25truein
\epsffile{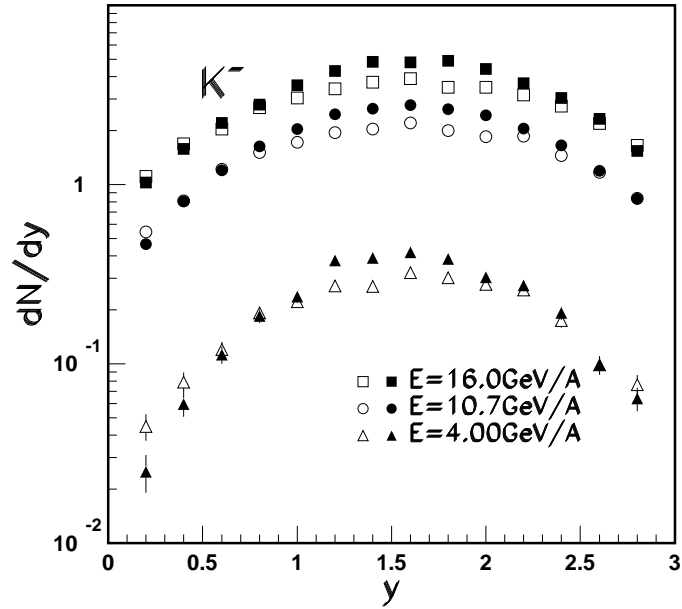}
\vskip 0.3truein
\caption{Same as Fig. 11 for the rapidity distribution of $K^-$ with
and without the mean field potential.}}
\end{figure}

\section{Summary}

In summary, we have studied $K^-$ production in Au+Au collisions at
different beam energies, based on an extension of the relativistic
transport model ART 1.0. Since $K^-$ has not been well treated in a
previous version of this model, we have improved the model by
including various reaction channels for both $K^-$ production and
scattering. Furthermore, an attractive $K^-$ mean field potential,
which is consistent with that extracted from the kaonic atom data,
has been included. Our results suggest that $K^-$ is mainly
produced from meson-meson and meson-baryon interactions. The
baryon-baryon interactions are less important, contributing only
one fifth of the total $K^-$ yield. Our model is able to describe
reasonably well the observed $K^-$ rapidity and transverse momentum
distributions. Furthermore, the mean field effect is much more
clearly seen on the $K^-$ rapidity distribution and transverse mass
spectrum, compared to those for $K^+$. Also, the $K^+/K^-$ ratio as
a function of transverse momentum or rapidity offers another
possibility for studying the medium effects.

We have also used the model to study the $K^-$ flow and found that
due to the attractive mean field potential the $K^-$ flow is
similar to the nucleon flow. Furthermore, the medium effect on
$K^-$ flow is much stronger than that on the $K^+$ flow, which was
previously studied using the ART 1.0 model.

We have compared the effect due to the nuclear potential to that
due to the Coulomb potential. Our results show that, unlike the
case of $K^+$ where the two effects are comparable, the nuclear
mean field potential has a much stronger effect on $K^-$ than the
Coulomb potential. However, both nuclear and Coulomb potentials
affect the $K^-$ in a similar way, i.e., they both tend to shift
the $K^-$ from high momentum to a lower value.

We have also carried out an extensive calculation for the $K^+$ and
$K^-$ yields, rapidity distributions and $m_t$ spectra at different
beam energies with and without mean field potentials. We have found
that for $K^-$ medium effects are almost independent of the
incident energy due to the fact that although $K^-$ produced at
lower energies are more susceptible to the influence of mean field
potential, which becomes, however, weaker as the beam energy is
reduced.

To conclude, it will be very useful to have more experimental data
from heavy ion collisions at the AGS energies to test the
predictions from our theoretical studies. Such a study will help
improve our understanding of medium effects on antikaons.

\bigskip
\centerline{\bf Acknowledgement}
\bigskip

We would like to thank C. Ogilvie for his interest in this study
and B. Zhang for a critical reading of the manuscript. This work
was supported in part by NSF Grant No. PHY-9509266 and PHY-9870038,
the Robert A Welch foundation under Grant A-1358, and the Texas
Advanced Research Program.

\end{document}